\documentclass[journal]{IEEEtran}
\usepackage{textgreek}

\usepackage{colortbl}    
\definecolor{blue}{RGB}{0,63,87}
\definecolor{colorcommentframe}{RGB}{0,112,155}
\usepackage{xcolor}
\usepackage{float}
\usepackage{caption} 
\usepackage{multirow} 
\usepackage{textgreek}
\usepackage{amssymb}
\usepackage{units}
\usepackage{orcidlink}
\usepackage{multirow}
\usepackage{array}
\usepackage{tabularx}
\usepackage{tikz}
\usetikzlibrary{calc,positioning}
\usepackage{lipsum,adjustbox}
\usepackage{colortbl}

\usepackage{cite}

\ifCLASSINFOpdf
\else
\fi
\usepackage{amsmath}
\usepackage{amssymb}
\usepackage{bm}
\usepackage{booktabs}
\usepackage{physics}
\usepackage{graphicx}
\usepackage{subcaption}

\begin{document}
%
\title{A Human-AI Teaming Framework for Deep Reinforcement Learning-Based Voltage Regulation in Distribution Networks}
%
%
%

\author{Mahmuda Akter,~\IEEEmembership{Student Member,~IEEE}, Hamidreza Nazaripouya,~\IEEEmembership{Senior Member,~IEEE}

}
\maketitle

 \begin{abstract}
The growing penetration of distributed energy resources (DERs) has increased the operational variability of distribution networks, making voltage regulation increasingly challenging. Conventional deep reinforcement learning (DRL) methods exhibit unsafe exploration behavior, slow convergence, and limited reliability, which restrict their applicability in safety-critical power system settings. This paper presents a human-interactive reinforcement learning framework that enhances the safety and robustness of autonomous voltage regulation. The proposed approach integrates a Soft Actor–Critic (SAC) agent with an adaptive Lagrange constraint mechanism to enforce voltage limits, while a human-guidance module provides sensitivity-based corrections through coordinated capacitor-bank and Battery Energy Storage System (BESS) dispatch. These corrections are incorporated into a human-regularized actor loss, enabling the policy to internalize safe and interpretable control behavior. The framework is implemented in the PowerGym–OpenDSS environment using the IEEE 13-node feeder. Simulation results show that the proposed Human-Interactive Lagrangian SAC (HI-LSAC) achieves significantly lower voltage-violation severity and reduced power losses compared with the other baseline methods.

\end{abstract}

%
\IEEEpeerreviewmaketitle

\section{Introduction}
\label{sec:intro}

\IEEEPARstart{T}{he} increasing penetration of renewable energy installations \cite{Powerlectronics-RES} and distributed energy resources (DERs) has introduced significant challenges to modern power grids, including difficulties in regulating voltage profiles and increased power losses in distribution networks. Voltage regulation refers to the coordinated management of voltage profiles within power distribution systems to ensure safe and efficient operation through intelligent dispatch of control devices such as voltage regulators, capacitor banks, smart inverters, and battery energy storage systems (BESS). The primary objective of voltage regulation is to maintain voltage profiles within acceptable limits (typically within $\pm 5\%$ of the nominal value) while minimizing network losses and enhancing overall power distribution efficiency \cite{distributedvolt/var,frameworkvolt/var}.

Physical model-driven optimization frameworks formulate the voltage regulation problem as an optimal power flow (OPF) problem \cite{sizingcapacitor}. These problems are typically solved using mathematical programming techniques such as linear programming, nonlinear programming, or mixed-integer nonlinear programming (MINLP) \cite{MINLP1}. While such approaches can achieve near-optimal solutions in simulation settings, they rely heavily on complete and accurate knowledge of network topology, equipment  \cite{parameters}, and system operating conditions. As system size and the number of distributed generators increase, these centralized optimization schemes face exponentially rising computational and communication burdens, leading to slower response times and reduced scalability.  References such as \cite{impactofDGsonvoltageregulation}, \cite{reviewofchallenges} and \cite{Voltageregulationsurvey} illustrate the applicability and limitations of centralized methods in smart grid scenarios.

Decentralized \cite{reviewcontrolstrategies} and distributed control strategies\cite{Distributeddecentralizedvoltagecontrol} are developed as scalable and robust alternatives to centralized Volt/Var control. These architectures reduce computational and communication burdens while enhancing resilience to failures. Reference \cite{Agentbaseddistributed} develops a fully distributed multi-agent Volt/Var control algorithm, where intelligent agents coordinate control devices to maintain voltage profiles and minimize network losses. Reference \cite{bi-levelvolt/var} proposes a bi-level Volt/Var optimization framework for conservation voltage reduction (CVR) that coordinates legacy voltage regulation devices with smart inverter control using mixed-integer linear and nonlinear OPF formulations. Based on the alternating direction method of multipliers (ADMM), a fully distributed second-order cone programming solver (D-SOCP) has been formulated in \cite{fulldistributed} for reactive power optimization. In \cite{combinedvolt/var}, a combined decentralized and local voltage regulation approach of soft open points has been proposed to regulate voltage profiles. However, these distributed control approaches still depend on precise network models, parameter accuracy, and complete system observability, which are difficult to obtain in real-world distribution networks characterized by frequent topology changes, measurement sparsity, and uncertainties in renewable generation. Additionally, the computational time of these physical model-based approaches grows exponentially with the size of the distribution network, making real-time implementation challenging in large-scale systems.

To address the limitations of model-based methods, Deep Reinforcement Learning (DRL) has recently been applied to voltage regulation problems \cite{RL3}. DRL combines the representation power of Deep Learning (DL) with the decision-making capability of Reinforcement Learning (RL), enabling agents to approximate complex system dynamics through neural networks while learning optimal control policies via interaction with the environment. A Deep Q-Network (DQN)-based Volt/Var optimization method for unbalanced distribution systems was proposed in \cite{RL4}, demonstrating the ability of DRL to manage nonlinear and unbalanced grid conditions. 

To coordinate devices with different response times, two-timescale voltage regulation strategies have been widely adopted. For example, \cite{RL7} proposes an RL framework that can improve safety during exploration, provide adaptive constraint enforcement, and enhance interpretability. Furthermore, Multi-Agent Reinforcement Learning (MARL) has gained increasing attention due to its distributed nature, allowing multiple agents to collaborate and optimize both local and global objectives. A MARL framework employing centralized training and decentralized execution (CTDE) was proposed in \cite{RL8} for voltage regulation, while a multi-agent Deep Deterministic Policy Gradient (DDPG)-based approach was developed in \cite{RL10} for autonomous voltage regulation. In addition, a communication-efficient consensus-based strategy was introduced in \cite{RL9} to enable individual agents to learn coordinated Volt/Var control policies using local rewards. \cite{RL7} also proposes a constrained Soft Actor–Critic (SAC)-based DRL method to address the online Volt/Var control problem under operational constraints.

While the above DRL and MARL-based strategies have demonstrated significant potential for autonomous voltage regulation, several challenges remain. Most existing approaches rely purely on data-driven exploration and lack explicit safety mechanisms during training, which can lead to unsafe voltage excursions and device stress in real-world deployment. Moreover, these methods often require extensive trial-and-error interactions with the environment to converge, resulting in slow learning and high computational costs. 

In addition, the majority of existing methods employ penalty-based reinforcement learning approaches to incorporate operational constraints. In such formulations, constraint violations are penalized through fixed weighting factors added to the reward function, thereby converting the constrained control problem into a reward-based optimization problem. However, the selection of an appropriate penalty weight is challenging. A small penalty weight may fail to adequately discourage unsafe actions, resulting in persistent voltage violations, whereas an excessively large penalty weight may make the agent overly conservative and restrict exploration. Moreover, ensuring safety only for the final learned policy is insufficient; unsafe exploration during training must also be mitigated to reduce voltage violations and promote stable learning.

These limitations highlight the need for an RL framework that can improve safety during exploration, adaptive constraint enforcement, and improved interpretability, which motivates the development of a human-interactive deep reinforcement learning approach that integrates human guidance directly into the training process to ensure secure and stable voltage regulation.

\section{Contributions}

The key contributions of this study are summarized as follows.

First, a human-interactive safe deep reinforcement learning (DRL) framework is developed for voltage regulation in distribution networks. Traditional DRL approaches often suffer from unsafe exploration and lack of interpretability, limiting their practical deployment in safety-critical grid environments. The framework integrates human guidance-based decision supervision to achieve safe and reliable voltage regulation. During training, a human-guidance module designed to emulate operator decision-making supervises DRL-generated actions and modifies them when voltage violations are predicted, thereby reducing unsafe exploration. By embedding operator knowledge into the decision-making loop, the agent learns constraint-compliant actions and achieves faster convergence and improved robustness compared to fully autonomous DRL methods.

Second, a sensitivity-based human-guided voltage correction mechanism is designed to emulate operator reasoning and provide interpretable corrective actions. The mechanism supervises and adjusts DRL-generated actions through coordinated control of heterogeneous devices, including capacitor banks and battery energy storage systems (BESS). Unlike existing methods that rely primarily on reactive power correction from a single device type such as PV inverters, the proposed approach coordinates both active- and reactive-power support through multi-device coordination. This physics-informed correction enhances learning safety and enables the agent to observe safe transitions that improve constraint compliance over successive training episodes.

Third, a Human-Interactive Lagrangian Soft Actor–Critic (HI-LSAC) algorithm is introduced to achieve safe and efficient voltage regulation in unbalanced distribution systems. The HI-LSAC framework extends the conventional LSAC by embedding human-guided decision supervision into the actor network and incorporating adaptive Lagrange-based constraint enforcement. In addition, a trend-based adjustment mechanism is introduced for the Lagrange multiplier to ensure smooth and stable updates of the penalty coefficient. This adaptive multiplier update dynamically strengthens or relaxes the constraint penalty based on the observed violation trends, preventing oscillations and improving training stability. The reformulated actor loss combines human-guidance and constraint terms, allowing the policy to learn physically interpretable and constraint-compliant control actions. This unified formulation enhances learning stability, accelerates convergence, and achieves reliable voltage regulation under time-varying operating conditions.

\section{Problem Formulation}
\label{sec:problem_formulation}

\subsection{Preliminaries of Constrained Markov Decision Process}
\label{sec:cmdp_prelim}

In reinforcement learning (RL), sequential decision-making problems are commonly modeled as a Markov decision process (MDP), defined by the tuple
\begin{equation}
\mathcal{M}
=
\left(
\mathcal{S},
\mathcal{A},
P,
r,
\gamma
\right),
\label{eq:mdp}
\end{equation}
where $\mathcal{S}$ denotes the state space, $\mathcal{A}$ denotes the action space, $P(s_{t+1}\mid s_t,a_t)$ represents the state-transition probability, $r(s_t,a_t,s_{t+1})$ is the immediate reward, and $\gamma\in(0,1]$ is the discount factor. The agent learns a stochastic policy $\pi_{\phi}(a_t\mid s_t)$, parameterized by $\phi$, that maximizes the expected cumulative discounted reward
\begin{equation}
J_r(\pi_{\phi})
=
\mathbb{E}_{\tau\sim\pi_{\phi}}
\left[
\sum_{t=0}^{T-1}
\gamma^{t}r_t
\right],
\label{eq:mdp_return}
\end{equation}
where $\tau$ denotes a trajectory generated by the policy.

However, in voltage regulation applications, the operating point of a distribution network must also satisfy strict voltage magnitude constraints at all energized buses and phases. This motivates the use of a constrained Markov decision process (CMDP) \cite{altmancmdp}, defined by
\begin{equation}
\mathcal{M}_{c}
=
\left(
\mathcal{S},
\mathcal{A},
P,
r,
c,
\gamma,
d_c
\right),
\label{eq:cmdp}
\end{equation}
where $c(s_t,a_t,s_{t+1})\geq0$ denotes the voltage-safety cost and $d_c$ represents the allowable expected cumulative cost. The CMDP objective is formulated as
\begin{align}
\max_{\pi_{\phi}}\quad
J_r(\pi_{\phi})
&=
\mathbb{E}_{\tau\sim\pi_{\phi}}
\left[
\sum_{t=0}^{T-1}
\gamma^{t}r_t
\right],
\label{eq:cmdp_reward}
\\
\mathrm{s.t.}\quad
J_c(\pi_{\phi})
&=
\mathbb{E}_{\tau\sim\pi_{\phi}}
\left[
\sum_{t=0}^{T-1}
\gamma^{t}c_t
\right]
\leq d_c.
\label{eq:cmdp_cost}
\end{align}

The constrained optimization problem can be represented through the Lagrangian
\begin{equation}
\mathcal{J}
\left(
\pi_{\phi},
\lambda_v
\right)
=
J_r(\pi_{\phi})
-
\lambda_v
\left[
J_c(\pi_{\phi})-d_c
\right],
\label{eq:cmdp_lagrangian}
\end{equation}
where $\lambda_v\geq0$ is the Lagrange multiplier associated with the voltage constraint. Instead of using a fixed penalty coefficient, $\lambda_v$ is adaptively updated during training according to the observed voltage-violation cost.

\subsection{Network Model and Power-Flow Constraints}
\label{sec:network_model}

We represent the power distribution system as a directed tree graph
$(\mathcal{N},\xi)$, where $\mathcal{N}$ is the set of nodes or buses and $\xi$ is the set of edges representing distribution lines and transformers. For each edge $(i,j)\in\xi$, node $i$ is defined as the parent of node $j$.

\begin{figure}[t]
    \centering
    \includegraphics[width=\linewidth]{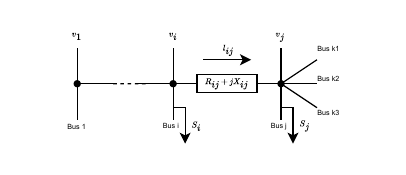}
    \caption{Branch-flow model of the radial distribution network.}
    \label{fig:branch_flow}
\end{figure}

For notational compactness, the branch-flow relationships are expressed for each energized phase $\varphi$. The physical power flow \cite{branchflowmodel} is governed by 
\begin{align}
p_{j,t}^{\varphi}
&=
p_{ij,t}^{\varphi}
-
R_{ij}^{\varphi}\ell_{ij,t}^{\varphi}
-
\sum_{(j,k)\in\xi}
p_{jk,t}^{\varphi},
\label{eq:branch_p}
\\
q_{j,t}^{\varphi}
&=
q_{ij,t}^{\varphi}
-
X_{ij}^{\varphi}\ell_{ij,t}^{\varphi}
-
\sum_{(j,k)\in\xi}
q_{jk,t}^{\varphi},
\label{eq:branch_q}
\\
\left(V_{j,t}^{\varphi}\right)^2
&=
\left(V_{i,t}^{\varphi}\right)^2
-
2\left(
R_{ij}^{\varphi}p_{ij,t}^{\varphi}
+
X_{ij}^{\varphi}q_{ij,t}^{\varphi}
\right)
\nonumber\\
&\quad+
\left[
\left(R_{ij}^{\varphi}\right)^2
+
\left(X_{ij}^{\varphi}\right)^2
\right]
\ell_{ij,t}^{\varphi},
\label{eq:branch_v}
\\
\ell_{ij,t}^{\varphi}
&=
\frac{
\left(p_{ij,t}^{\varphi}\right)^2
+
\left(q_{ij,t}^{\varphi}\right)^2
}{
\left(V_{i,t}^{\varphi}\right)^2
}.
\label{eq:branch_i}
\end{align}

Here, $p_{ij,t}^{\varphi}$ and $q_{ij,t}^{\varphi}$ denote the active and reactive power flows from node $i$ to node $j$, respectively, whereas $p_{j,t}^{\varphi}$ and $q_{j,t}^{\varphi}$ represent the net active and reactive power demand at node $j$. The variable $V_{i,t}^{\varphi}$ denotes the voltage magnitude and $\ell_{ij,t}^{\varphi}$ represents the squared current magnitude of branch $(i,j)$. The parameters $R_{ij}^{\varphi}$ and $X_{ij}^{\varphi}$ denote the corresponding branch resistance and reactance.

The above relationships illustrate the nonlinear and nonconvex nature of the voltage regulation problem. In the implemented environment, the complete unbalanced three-phase power-flow equations and detailed device models are solved using OpenDSS.

The total active power loss is expressed as
\begin{equation}
P_{\mathrm{loss},t}
=
\sum_{(i,j)\in\xi}
\sum_{\varphi\in\Phi_{ij}}
R_{ij}^{\varphi}
\ell_{ij,t}^{\varphi},
\label{eq:power_loss}
\end{equation}
where $\Phi_{ij}$ denotes the set of energized phases on branch $(i,j)$.

The network must satisfy the voltage limits
\begin{equation}
V_{\min}
\leq
V_{i,t}^{\varphi}
\leq
V_{\max},
\quad
\forall i\in\mathcal{N},
\quad
\varphi\in\Phi_i.
\label{eq:voltage_limits}
\end{equation}

The capacitor-bank switching states satisfy
\begin{equation}
x_{k,t}^{\mathrm{cap}}
\in
\{0,1\},
\quad
\forall k\in\mathcal{N}_{\mathrm{cap}}.
\label{eq:cap_limits}
\end{equation}

For each BESS $b\in\mathcal{N}_{\mathrm{bat}}$, the active-power, reactive-power, and apparent-power constraints are
\begin{align}
P_b^{\min}
&\leq
P_{b,t}^{\mathrm{bat}}
\leq
P_b^{\max},
\label{eq:bess_p_limits}
\\
Q_b^{\min}
&\leq
Q_{b,t}^{\mathrm{bat}}
\leq
Q_b^{\max},
\label{eq:bess_q_limits}
\\
\left(P_{b,t}^{\mathrm{bat}}\right)^2
+
\left(Q_{b,t}^{\mathrm{bat}}\right)^2
&\leq
\left(S_b^{\max}\right)^2.
\label{eq:bess_s_limits}
\end{align}

The BESS state of charge is constrained by
\begin{equation}
\mathrm{SOC}_{b}^{\min}
\leq
\mathrm{SOC}_{b,t}
\leq
\mathrm{SOC}_{b}^{\max}.
\label{eq:bess_soc_limits}
\end{equation}

\subsection{Formulation of Voltage Regulation Optimization}
\label{sec:voltage_optimization}

The primary objective of voltage regulation is to improve distribution-network operating efficiency while maintaining bus voltages and controllable devices within their prescribed operating limits. Accordingly, the operating objective considered in this work is expressed as

\begin{equation}
\min_{\{a_t\}_{t=0}^{T-1}}
\sum_{t=0}^{T-1}
\left(
w_{\ell}\eta_{\mathrm{loss},t}
+w_{\mathrm{cap}}C_{\mathrm{cap},t}
+w_{\mathrm{dis}}C_{\mathrm{dis},t}
+w_{\mathrm{soc}}C_{\mathrm{soc},t}
\right).
\label{eq:operating_objective}
\end{equation}

Here,
\begin{equation}
\eta_{\mathrm{loss},t}
=
\frac{
P_{\mathrm{loss},t}
}{
P_{\mathrm{total},t}
}
\label{eq:loss_ratio}
\end{equation}
denotes the normalized active power loss. The term
$C_{\mathrm{cap},t}$ penalizes unnecessary capacitor-bank switching,
$C_{\mathrm{dis},t}$ discourages excessive BESS discharge, and
$C_{\mathrm{soc},t}$ represents the terminal deviation of the BESS
state of charge from its desired value. The device operating limits are not incorporated into the objective
as soft penalty terms. Instead, the admissible capacitor and BESS
actions are restricted by the corresponding device operating limits,
while the resulting system operating point is obtained through the
distribution-network state transition. Voltage-limit violations are
handled separately through the CMDP safety cost and the adaptive
Lagrangian mechanism introduced in the subsequent sections.

Rather than solving \eqref{eq:operating_objective} repeatedly using an
explicit optimization model, the proposed HI-LSAC framework learns
a control policy through interactions with the distribution-system
environment. Human guidance is further introduced during training
to identify and correct unsafe policy actions before they are
committed to the environment.

\subsection{Voltage Regulation as a CMDP}
\label{sec:voltage_cmdp}

The voltage regulation task is formulated as the CMDP introduced in Section~\ref{sec:cmdp_prelim}. At each decision interval $t$, the agent observes the current system state $s_t\in\mathcal{S}$ and generates a policy action $a_t^{\pi}\in\mathcal{A}$. Before the proposed action is committed to the environment, a non-committing OpenDSS power-flow simulation evaluates its expected post-action voltage response. Based on the previewed operating condition, the action is either directly accepted or modified through the proposed human-guided correction mechanism.

\subsubsection{State Definition}

The system state at time $t$ contains the three-phase bus voltage magnitudes, capacitor-bank switching states, BESS operating conditions, and available exogenous operating information:
\begin{equation}
s_t
=
\left[
\mathbf{V}_t,
\mathbf{x}_{t}^{\mathrm{cap}},
\mathbf{SOC}_t,
\mathbf{P}_{t}^{\mathrm{bat}},
\mathbf{Q}_{t}^{\mathrm{bat}},
\bm{\zeta}_t
\right],
\label{eq:state}
\end{equation}
where
\begin{equation}
\mathbf{V}_t
=
\left\{
V_{i,t}^{\varphi}
\mid
i\in\mathcal{N},
\;
\varphi\in\Phi_i
\right\}.
\end{equation}

Here, $\mathbf{x}_{t}^{\mathrm{cap}}$ contains the capacitor-bank states; $\mathbf{SOC}_t$, $\mathbf{P}_{t}^{\mathrm{bat}}$, and $\mathbf{Q}_{t}^{\mathrm{bat}}$ describe the BESS operating states; and $\bm{\zeta}_t$ represents available exogenous information such as the load profile and, when applicable, PV generation.

\subsubsection{Action Definition}

The physical control action is expressed as
\begin{equation}
a_t
=
\begin{bmatrix}
\mathbf{x}_{t}^{\mathrm{cap}} &
\mathbf{P}_{t}^{\mathrm{bat}} &
\mathbf{Q}_{t}^{\mathrm{bat}}
\end{bmatrix}^{\top}.
\label{eq:action}
\end{equation}

The stochastic SAC actor generates a normalized continuous action $a_t^{\pi}$. A deterministic action-mapping function converts the capacitor-bank components into binary switching commands and scales the BESS components to their corresponding active- and reactive-power operating ranges. Therefore, the proposed policy can simultaneously coordinate discrete and continuous voltage-control resources.

\subsubsection{Operating Reward}

The environmental reward represents the operating objective and is defined independently of the voltage-safety cost as
\begin{align}
r_t
=
-\Big(
&w_{\ell}\eta_{\mathrm{loss},t}
+
w_{\mathrm{cap}}C_{\mathrm{cap},t}
\nonumber\\
&+
w_{\mathrm{dis}}C_{\mathrm{dis},t}
+
w_{\mathrm{soc}}C_{\mathrm{soc},t}
\Big).
\label{eq:operating_reward}
\end{align}

The capacitor switching cost is expressed as
\begin{equation}
C_{\mathrm{cap},t}
=
\sum_{k\in\mathcal{N}_{\mathrm{cap}}}
\left|
x_{k,t}^{\mathrm{cap}}
-
x_{k,t-1}^{\mathrm{cap}}
\right|.
\label{eq:cap_cost}
\end{equation}

The term $C_{\mathrm{dis},t}$ discourages excessive BESS discharge, while $C_{\mathrm{soc},t}$ penalizes the terminal deviation of the BESS state of charge from its desired value.

\subsection{Human-Guided Action Correction}
\label{sec:human_correction}

To ensure safe and reliable voltage regulation during training, the proposed framework introduces a human-guided correction mechanism that supervises the actions generated by the DRL policy. At each control interval, the policy action $a_t^{\pi}$ is first evaluated through a non-committing power-flow simulation. Human guidance is activated when the previewed action results in at least one voltage violation:
\begin{equation}
I_t
=
\mathbb{I}
\left[
\exists(i,\varphi):
\widehat{V}_{i,t+1}^{\varphi}
\notin
[V_{\min},V_{\max}]
\right].
\label{eq:intervention}
\end{equation}

For each violated bus phase, the signed voltage deviation is calculated as
\begin{equation}
\Delta V_{i,t}^{\varphi}
=
\begin{cases}
V_{\min}
-
\widehat{V}_{i,t+1}^{\varphi},
&
\widehat{V}_{i,t+1}^{\varphi}<V_{\min},
\\[3pt]
V_{\max}
-
\widehat{V}_{i,t+1}^{\varphi},
&
\widehat{V}_{i,t+1}^{\varphi}>V_{\max},
\\[3pt]
0,
&
\text{otherwise}.
\end{cases}
\label{eq:signed_voltage_violation}
\end{equation}

According to \eqref{eq:signed_voltage_violation}, $\Delta V_{i,t}^{\varphi}>0$ represents an undervoltage condition, whereas $\Delta V_{i,t}^{\varphi}<0$ represents an overvoltage condition.

\subsubsection{Capacitor-Bank Correction}

When a controllable capacitor bank is available at the violated bus phase, both feasible switching states are evaluated using non-committing power-flow simulations. The switching state producing the smaller remaining voltage violation is selected as the candidate human action:
\begin{equation}
x_{k,t}^{\mathrm{H}}
=
\arg\min_{x\in\{0,1\}}
\Gamma
\left[
\mathcal{F}_{V}
\left(
s_t,
a_t^{\pi}[x_k\leftarrow x]
\right)
\right],
\label{eq:cap_correction}
\end{equation}
where $\Gamma(\cdot)$ denotes the corresponding voltage-violation measure. The capacitor action is accepted only when it improves the voltage condition relative to the current candidate action.

\subsubsection{BESS Reactive-Power Correction}

When capacitor-bank switching is unavailable or insufficient, BESS reactive power is adjusted to mitigate the remaining voltage deviation. At correction iteration $m$, the reactive-power command is updated as
\begin{equation}
Q_{b,t}^{\mathrm{H},(m+1)}
=
\Pi_{\mathcal{Q}_b}
\left[
Q_{b,t}^{\mathrm{H},(m)}
+
\frac{
\Delta V_{i,t}^{\varphi,(m)}
}{
S_{i,b,t}^{VQ,\varphi}
}
\right],
\label{eq:q_correction}
\end{equation}
where $\Pi_{\mathcal{Q}_b}$ represents projection onto the feasible reactive-power operating range of BESS $b$.

The voltage--reactive-power sensitivity is estimated using central perturbations:
\begin{equation}
S_{i,b,t}^{VQ,\varphi}
\approx
\frac{
V_{i,t+1}^{\varphi}
\left(
Q_{b,t}+\delta Q_b
\right)
-
V_{i,t+1}^{\varphi}
\left(
Q_{b,t}-\delta Q_b
\right)
}{
2\delta Q_b
}.
\label{eq:vq_sensitivity}
\end{equation}

This sensitivity quantifies the voltage response at bus $i$ and phase $\varphi$ to a change in the BESS reactive-power command.

\subsubsection{BESS Active-Power Correction}

If a residual voltage violation remains after reactive-power correction, the BESS active-power command is updated as
\begin{equation}
P_{b,t}^{\mathrm{H},(m+1)}
=
\Pi_{\mathcal{P}_b}
\left[
P_{b,t}^{\mathrm{H},(m)}
+
\frac{
\Delta V_{i,t}^{\varphi,(m)}
}{
S_{i,b,t}^{VP,\varphi}
}
\right],
\label{eq:p_correction}
\end{equation}
where $\Pi_{\mathcal{P}_b}$ denotes projection onto the feasible active-power operating range. The voltage--active-power sensitivity is estimated as
\begin{equation}
S_{i,b,t}^{VP,\varphi}
\approx
\frac{
V_{i,t+1}^{\varphi}
\left(
P_{b,t}+\delta P_b
\right)
-
V_{i,t+1}^{\varphi}
\left(
P_{b,t}-\delta P_b
\right)
}{
2\delta P_b
}.
\label{eq:vp_sensitivity}
\end{equation}

The final BESS commands are subsequently projected onto the physical feasibility region defined by \eqref{eq:bess_p_limits}--\eqref{eq:bess_soc_limits}. This step leverages the coupling between active power injections and local voltages to provide additional voltage support when reactive power correction alone is insufficient. When the local correction sequence remains insufficient, additional feeder-wide candidate actions are evaluated, and the action that provides the greatest reduction in the overall voltage-violation severity is retained.

The complete human-guided corrective action is expressed as
\begin{equation}
\begin{aligned}
a_t^{\mathrm{H}}
&=
\begin{bmatrix}
a_{t,\mathrm{CB}}^{\mathrm{H}} \\
a_{t,\mathrm{BESS}}^{\mathrm{H}}
\end{bmatrix},
\qquad
a_{t,\mathrm{CB}}^{\mathrm{H}}
\in \{0,1\}^{N_{\mathrm{CB}}},
\\
a_{t,\mathrm{BESS}}^{\mathrm{H}}
&=
\begin{bmatrix}
P_t^{\mathrm{bat,H}} &
Q_t^{\mathrm{bat,H}}
\end{bmatrix}^{\top}.
\end{aligned}
\label{eq:human_action_def_final}
\end{equation}

The correction procedure is repeated until the previewed voltage violations are eliminated or the maximum number of correction iterations is reached. The final control action is selected as

\begin{equation}
a_t
=
\begin{cases}
a_t^{\text{H}},
&
\text{if human correction is triggered},
\\[4pt]
a_t^{\pi},
&
\text{otherwise}.
\end{cases}
\label{eq:final_control_action}
\end{equation}

\subsection{Proposed Human-Interactive Lagrangian Soft Actor--Critic Algorithm}
\label{sec:HI-LSAC}

To achieve safe and adaptive voltage regulation in distribution networks, we propose the Human-Interactive Lagrangian Soft Actor--Critic (HI-LSAC) algorithm. The proposed method extends the standard SAC framework by integrating 1) adaptive Lagrangian constraint learning for voltage regulation, 2) human-guided actor regularization that directs the learned policy toward corrected actions, and 3) learning from rejected unsafe actions and corresponding human-corrected actions. These mechanisms allow the agent to improve voltage safety while preserving the sample efficiency and exploration capability of SAC.

\subsubsection{Overview of SAC Framework}

The SAC algorithm employs an actor--critic architecture consisting of two critic networks, $Q_{\theta_1}$ and $Q_{\theta_2}$, their corresponding target networks, and a stochastic actor network $\pi_{\phi}(a|s)$. The use of twin critic networks mitigates overestimation bias by retaining the minimum of the two estimated Q-values during policy learning.

To incorporate human-guided actions, the control action at time step $t$ is determined by
\begin{equation}
a_t
=
I_t a_t^{\mathrm{H}}
+
(1-I_t)a_t^{\pi},
\label{eq:hl_action}
\end{equation}
where $a_t^{\mathrm{H}}$ is the human-corrected action, $a_t^{\pi}$ is the action generated by the policy network, and $I_t\in\{0,1\}$ denotes the human-intervention indicator. When the previewed policy action causes a voltage violation, $I_t=1$ and the human-corrected action is applied; otherwise, $I_t=0$ and the original policy action is retained.

The SAC algorithm seeks to maximize the expected return while maintaining policy entropy to encourage sufficient exploration. Its maximum-entropy objective is expressed as
\begin{equation}
\max_{\pi}
\mathbb{E}
\left[
\sum_{t=0}^{T-1}
\gamma^t
\left(
r_t
+
\alpha
\mathcal{H}
\left(
\pi(\cdot|s_t)
\right)
\right)
\right].
\label{eq:sac_objective}
\end{equation}
where $\alpha$ denotes the entropy-temperature parameter.

\subsubsection{Lagrangian Reward Shaping and Critic Update}

Although SAC provides efficient off-policy learning, it does not explicitly account for voltage constraints. In this work, the voltage-safety cost $c_t$ represents the aggregate magnitude of bus-phase voltage-limit violations obtained from the non-committing power-flow simulation. Thus, $c_t=0$ when all bus-phase voltages remain within the prescribed limits, whereas $c_t>0$ indicates the presence and severity of voltage violations.

The constraint cost is first normalized as
\begin{equation}
\bar{c}_t
=
\frac{c_t}{\sigma_c},
\label{eq:normalized_constraint}
\end{equation}
where $\sigma_c$ is a scaling parameter. The reward used for critic learning is then defined as
\begin{equation}
\widetilde{r}_t
=
r_t
-
\lambda_v\bar{c}_t.
\label{eq:lagrange_reward}
\end{equation}

Unlike conventional penalty-based approaches that use a fixed voltage-violation penalty, $\lambda_v$ is adaptively updated during training according to the observed constraint violations.

For the next state $s_{t+1}$, an action $a_{t+1}$ is sampled from the stochastic policy, and the soft Bellman target is calculated as
\begin{align}
y_t
=
\widetilde{r}_t
+
\gamma(1-d_t)
\Big[
&
\min_{j\in\{1,2\}}
Q_{\bar{\theta}_j}
(s_{t+1},a_{t+1})
\nonumber\\
&-
\alpha
\log
\pi_{\phi}
(a_{t+1}|s_{t+1})
\Big],
\label{eq:critic_target}
\end{align}
where $d_t$ denotes the terminal indicator.

The twin critic networks are optimized using
\begin{equation}
\mathcal{L}_{Q}
=
\frac{1}{2}
\sum_{j=1}^{2}
\mathbb{E}_{\mathcal{D}}
\left[
\left(
Q_{\theta_j}(s_t,a_t)
-
y_t
\right)^2
\right].
\label{eq:critic_loss}
\end{equation}

The minimum of the twin target Q-values reduces overestimation bias and improves training stability.

\subsubsection{Adaptive Lagrange Multiplier}

The Lagrange multiplier is updated according to the observed voltage-constraint cost using
\begin{equation}
\lambda_v
\leftarrow
\Pi_{[0,\lambda_{\max}]}
\left[
\lambda_v
+
\eta_{\lambda}
\left(
\widehat{\mathbb{E}}
[\bar{c}_t]
-
\epsilon_v
\right)
\right],
\label{eq:lambda_update}
\end{equation}
where $\eta_{\lambda}$ denotes the dual learning rate, $\epsilon_v$ is the allowable constraint threshold, and $\lambda_{\max}$ limits the maximum value of the multiplier.

When voltage violations persist, $\lambda_v$ increases and strengthens the voltage-constraint penalty in \eqref{eq:lagrange_reward}. Conversely, when the constraint cost remains close to the desired level, the growth of $\lambda_v$ is reduced. In this way, the relative importance of voltage safety is adjusted adaptively without requiring a manually selected fixed penalty coefficient.

\subsubsection{Human-Guided Actor Update}

The conventional SAC actor is optimized by maximizing the expected Q-value while maintaining sufficient policy entropy. The corresponding actor loss is
\begin{equation}
\mathcal{L}_{\pi}^{\mathrm{SAC}}
=
\mathbb{E}_{\substack{
s_t\sim\mathcal{D}\\
a_t\sim\pi_{\phi}
}}
\left[
\alpha
\log
\pi_{\phi}(a_t|s_t)
-
\min_{j\in\{1,2\}}
Q_{\theta_j}(s_t,a_t)
\right].
\label{eq:sac_actor_loss}
\end{equation}

To incorporate human guidance, the proposed method additionally directs the deterministic mean action of the stochastic policy toward the corresponding human-corrected action. Let $m_{\phi}(s_t)$ denote the pre-squashing Gaussian mean generated by the actor. The deterministic mean action is defined as
\begin{equation}
\mu_{\phi}(s_t)
=
\tanh
\left(
m_{\phi}(s_t)
\right).
\label{eq:actor_mean}
\end{equation}

For transitions containing human corrections, the guidance loss is defined as
\begin{equation}
\mathcal{L}_{\mathrm{H}}
=
\mathbb{E}_{\mathcal{D}_{\mathrm{H}}}
\left[
\left\|
\mu_{\phi}(s_t)
-
a_t^{\mathrm{H}}
\right\|_2^2
\right],
\label{eq:human_guidance_loss}
\end{equation}
where $\mathcal{D}_{\mathrm{H}}$ represents the human-guided samples in the replay minibatch.

The complete actor loss is therefore expressed as
\begin{equation}
\mathcal{L}_{\pi}
=
\mathcal{L}_{\pi}^{\mathrm{SAC}}
+
\omega_{\mathrm{H}}
\mathcal{L}_{\mathrm{H}},
\label{eq:HIlsac_actor_loss}
\end{equation}
where $\omega_{\mathrm{H}}$ controls the influence of human guidance.

The human-guidance term encourages the policy to imitate the corrective actions provided during unsafe operating conditions, while the standard SAC objective preserves autonomous policy optimization and exploration.

\subsubsection{Learning from Rejected Actions}

When a policy action $a_t^{\pi}$ is identified as unsafe because of a previewed voltage violation, both the rejected unsafe transition and the corresponding human-corrected transition are retained for learning. To further encourage the policy to imitate the human-corrected action, the rejected transition is assigned
\begin{align}
r_t^{\mathrm{rej}}
=
\min
\Big\{
&r_t^{\pi}
-
\kappa_{\mathrm{H}}
-
\kappa_{\mathrm{sev}}c_t,
\nonumber\\
&r_t^{\mathrm{H}}
-
\Delta_r
\Big\},
\label{eq:rejected_reward}
\end{align}
where $r_t^{\pi}$ and $r_t^{\mathrm{H}}$ denote the rewards associated with the rejected and human-corrected actions, respectively; $\kappa_{\mathrm{H}}$ denotes the intervention penalty; $\kappa_{\mathrm{sev}}$ weights the voltage-violation severity; and $\Delta_r>0$ specifies the minimum reward gap.

Accordingly,
\begin{equation}
r_t^{\mathrm{rej}}
\leq
r_t^{\mathrm{H}}
-
\Delta_r,
\label{eq:reward_gap_condition}
\end{equation}
providing the critic with a more favorable learning signal for the human-corrected action than for the corresponding rejected unsafe action.

\subsubsection{Replay Buffer Structure}

The replay buffer stores both ordinary DRL transitions and human-guided experiences. When the policy action does not trigger human correction, the stored transition is
\begin{equation}
\left(
s_t,
a_t^{\pi},
r_t,
c_t,
s_{t+1},
I_t=0
\right).
\label{eq:normal_replay}
\end{equation}

When human correction is triggered, both the rejected policy transition and the corrected transition are retained:
\begin{align}
\mathcal{D}
\leftarrow
\mathcal{D}
\cup
\Big\{
&
(
s_t,
a_t^{\pi},
r_t^{\mathrm{rej}},
c_t^{\pi},
\widehat{s}_{t+1}^{\pi},
I_t=0
),
\nonumber\\
&
(
s_t,
a_t^{\mathrm{H}},
r_t^{\mathrm{H}},
c_t^{\mathrm{H}},
s_{t+1},
I_t=1
)
\Big\}.
\label{eq:hybrid_replay}
\end{align}

The rejected transition allows the critic to learn the consequence of the unsafe action, whereas the corrected transition provides the corresponding safer control decision. During minibatch sampling, human-guided transitions are sampled together with ordinary DRL experiences so that informative corrective samples remain sufficiently represented during training. In addition, successfully corrected actions are stored in a separate expert
correction buffer,
\begin{equation}
\mathcal{D}_{\mathrm{E}}
=
\left\{
(s_t,a_t^{\mathrm{H}})
\right\}.
\label{eq:expert_buffer}
\end{equation}

The stored expert samples are periodically used to further guide the actor toward the human-corrected actions. The corresponding auxiliary
expert-guidance loss is defined as
\begin{equation}
\mathcal{L}_{\mathrm{E}}(\phi)
=
\beta_{\mathrm{E}}(k)
\mathbb{E}_{(s_t,a_t^{\mathrm{H}})\sim\mathcal{D}_{\mathrm{E}}}
\left[
\ell_{\mathrm{SL1}}
\left(
\mu_{\phi}(s_t),
a_t^{\mathrm{H}}
\right)
\right],
\label{eq:expert_guidance_loss}
\end{equation}
where $\ell_{\mathrm{SL1}}(\cdot)$ denotes the Smooth-$L_1$ loss and
$\beta_{\mathrm{E}}(k)$ is a decaying expert-guidance coefficient at
training update $k$. 

The coefficient is gradually reduced during training, allowing stronger supervision during the early learning stage while progressively increasing policy autonomy.

Overall, the proposed HI-LSAC framework integrates adaptive voltage-constraint learning, human-guided policy regularization, and learning from rejected and corrected actions within a unified off-policy framework. The adaptive Lagrangian mechanism regulates the influence of voltage violations, while the human-guided learning mechanisms enable the policy to gradually internalize corrective control behavior and reduce its dependence on human intervention.

\begin{figure*}[!t]
    \centering
    \includegraphics[width=\textwidth]{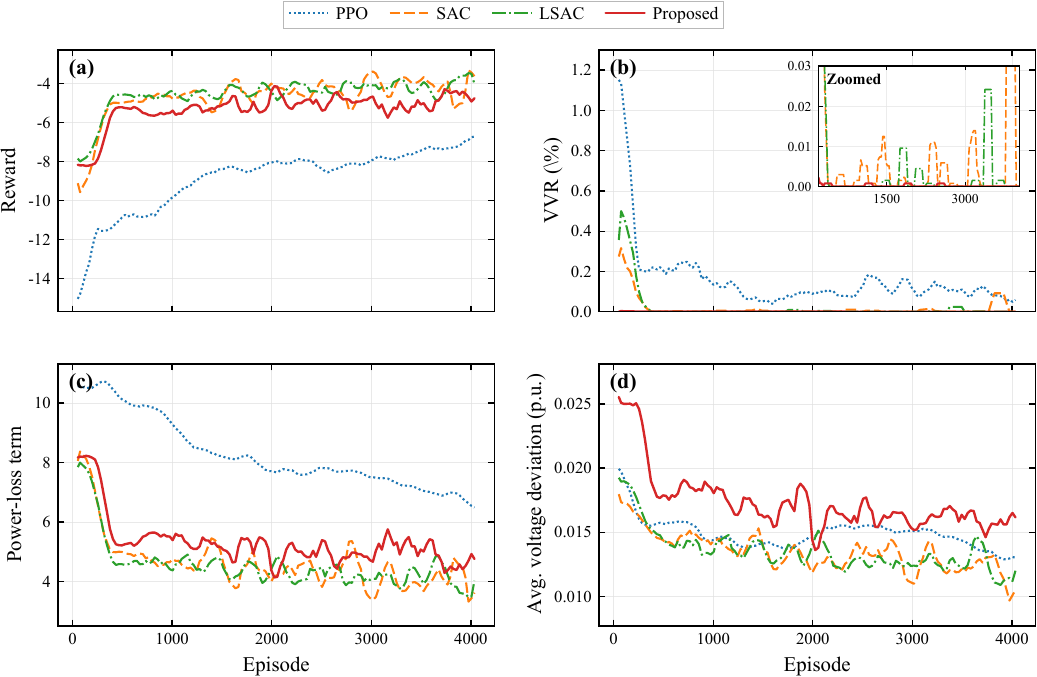}
    \caption{Training performance comparison of the investigated methods: (a) reward, (b) voltage violation ratio (VVR), (c) power loss, and (d) average voltage deviation.}
    \label{fig:training_performance}
\end{figure*}

\section{Case Study and Numerical Results}

The proposed HI-LSAC framework is evaluated on a modified IEEE 13-node test feeder implemented in the PowerGym environment, with power-flow simulations performed using OpenDSS. The test system incorporates time-varying load and PV generation profiles to represent daily operating variability. Capacitor banks and BESS active and reactive power are coordinated for voltage regulation and power-loss reduction. Each episode consists of 24 operating intervals, and the acceptable voltage range is set to $[0.95,\,1.05]$ p.u. The proposed framework is implemented in Python using PyTorch and Stable-Baselines3. All simulations and DRL training are conducted on a computer equipped with an Intel Core Ultra 7 155H processor, 64~GB of RAM, and an NVIDIA RTX 500 Ada Generation Laptop GPU with 4~GB of dedicated memory.

The modified IEEE 13-bus feeder includes distributed PV generation,
capacitor banks, and BESS resources. The PV generation is modeled as
an uncontrollable time-varying power injection, whereas the capacitor
banks and BESS units constitute the controllable voltage-regulation
resources. The capacitor banks are operated through binary switching
commands, while the BESS resources provide controllable active- and
reactive-power support.

\subsection{Training Performance}
\label{subsec:training_performance}

To evaluate the contribution of the proposed HI-LSAC framework, three
representative DRL methods are considered as baselines. The first baseline
is proximal policy optimization (PPO), a widely used on-policy
actor--critic algorithm that has also been adopted for Volt/Var control
and human-in-the-loop voltage-regulation studies
\cite{PPO}. PPO is included to provide a representative
on-policy DRL benchmark. The second baseline is the standard SAC
algorithm, which has been successfully applied to distribution-system
voltage control because of its off-policy learning and entropy-regularized
exploration \cite{SAC}. Since SAC also serves as the underlying
learning architecture of the proposed method, this comparison directly
demonstrates the effect of introducing constraint-aware and human-guided
learning mechanisms. The third baseline is Lagrangian SAC (LSAC), which
incorporates voltage constraints through adaptive Lagrangian learning but
does not employ the proposed human-guided action correction, actor
guidance, or learning from rejected actions. Safe off-policy SAC-based
methods have previously been investigated for Volt/Var control
\cite{sizingcapacitor}; therefore, LSAC provides a direct benchmark for
evaluating the additional contribution of human guidance beyond
Lagrangian constraint enforcement alone.

Fig.~\ref{fig:training_performance} compares the training performance of
PPO, SAC, LSAC, and the proposed HI-LSAC in terms of episode reward,
voltage violation ratio (VVR), power-loss term, and average voltage
deviation. These metrics are considered jointly because operating
efficiency and voltage security are not necessarily improved
simultaneously. In particular, a policy achieving a higher reward or a
smaller average voltage deviation does not necessarily produce fewer
voltage-limit violations.

As shown in Fig.~\ref{fig:training_performance}(a), all four methods
improve their episode rewards as training progresses. PPO exhibits the
slowest improvement and remains at a substantially lower reward level
throughout training. In contrast, the off-policy SAC-based methods improve
more rapidly because previously collected experiences can be repeatedly
used for policy updates. SAC and LSAC eventually attain the highest
training rewards, while HI-LSAC converges to a slightly lower reward
level.

The moderately lower training reward of HI-LSAC results from the
additional emphasis placed on voltage safety during exploration. Unlike
standard SAC, unsafe policy actions are not simply executed and reinforced
according to their immediate operating reward. When an unsafe action is
identified, the human-guided mechanism provides a corrected action, while
the rejected action is retained with a lower learning target through the
reward-gap mechanism. In addition, the adaptive Lagrangian penalty,
human-guided actor regularization, and expert-guided updates continuously
direct the learned policy toward actions that are more consistent with
safe operation. Consequently, the proposed agent sacrifices some
short-term reward during training rather than repeatedly exploiting
actions that may improve the operating objective at the expense of
voltage security.

The effectiveness of this learning mechanism is more clearly demonstrated
by the VVR results in Fig.~\ref{fig:training_performance}(b). Following
the definition in \cite{chen2025robust}, VVR is calculated as the
percentage of evaluated bus-phase voltage points that violate the
prescribed voltage limits:
\begin{equation}
\mathrm{VVR}
=
\frac{\sum_{t=1}^{N_{\mathrm{eval}}} n_{\mathrm{VV},t}}
{\sum_{t=1}^{N_{\mathrm{eval}}} n_{\mathrm{ph},t}}
\times 100\%,
\label{eq:vvr}
\end{equation}
where $n_{\mathrm{VV},t}$ denotes the number of bus-phase voltages outside
$[V_{\min},V_{\max}]$ at control interval $t$, $n_{\mathrm{ph},t}$ is the
total number of evaluated bus-phase voltage points, and $N_{\mathrm{eval}}$
is the total number of evaluated control intervals.

PPO initially exhibits the highest VVR and continues to experience
noticeable violations throughout the training process. SAC reduces the
VVR considerably faster, but intermittent violations remain because
voltage security is learned only indirectly from the reward signal. LSAC
further reduces these violations by adaptively increasing the influence
of the voltage-constraint cost. Nevertheless, as shown more clearly in
the enlarged inset, occasional violations still occur during the later
training stage.

In comparison, HI-LSAC rapidly reduces the VVR to nearly zero and
maintains the most consistent voltage-security performance throughout the
remaining training process. This improvement results from the
complementary roles of the proposed learning mechanisms. The Lagrangian
component provides a global constraint-learning signal, while the
human-guided correction mechanism acts directly when an unsafe action is
identified. The corrected action prevents the corresponding unsafe action
from being committed to the environment, and the paired rejected and
corrected transitions provide the critic with explicit information about
their relative quality. Meanwhile, human-guided actor regularization and
the expert correction buffer transfer this corrective behavior to the
policy itself. Therefore, the policy does not rely solely on repeated
trial-and-error interactions to discover the consequences of unsafe
actions.

Fig.~\ref{fig:training_performance}(c) presents the corresponding
power-loss term. PPO maintains the highest value throughout training,
consistent with its slower improvement in the operating objective. SAC
and LSAC attain the lowest converged power-loss terms, whereas HI-LSAC
converges to a slightly higher level. This difference reflects the
safety--efficiency tradeoff observed during training. SAC and LSAC retain
greater freedom to explore actions that reduce instantaneous losses,
including actions that occasionally lead to voltage-limit violations.
In contrast, HI-LSAC modifies such actions when violations are predicted
and directs the policy toward the corresponding corrected operating
points. The resulting power-loss term is therefore slightly higher during
training, but remains substantially lower than that of PPO while achieving
a markedly lower VVR.

The average voltage deviation shown in
Fig.~\ref{fig:training_performance}(d) provides a complementary measure of
voltage-regulation performance. SAC and LSAC achieve the lowest average
deviations after convergence, whereas HI-LSAC maintains a somewhat higher
value. This result is not inconsistent with its near-zero VVR. Average
voltage deviation measures the mean distance of all bus-phase voltage
magnitudes from the nominal value of $1.0$~p.u., whereas VVR directly
measures violations of the prescribed operating limits. The proposed
method is designed primarily to maintain voltage feasibility rather than
force every bus voltage toward exactly $1.0$~p.u. Therefore, voltages may
remain safely within the allowable range while being farther from the
nominal value, allowing unnecessary corrective actions to be avoided.
Average voltage deviation should thus be interpreted together with VVR
rather than as an independent measure of voltage security.

It should be noted that the VVR reported in
Fig.~\ref{fig:training_performance} characterizes the voltage security of
the actions actually executed during training, including human-corrected
actions. Therefore, these results demonstrate the effectiveness of the proposed framework in reducing unsafe exploration rather than, by themselves, establishing the autonomous safety of the learned policy. The
policy-only performance of the trained agents is consequently evaluated
in the following subsection without online human intervention.

\subsection{Evaluation}

\begin{table}[t]
\centering
\caption{PERFORMANCE COMPARISON OF DIFFERENT VOLTAGE CONTROL METHODS IN ONLINE APPLICATION}
\label{tab:final_eval_powerloss_vvr}
\small
\setlength{\tabcolsep}{4.5pt}
\renewcommand{\arraystretch}{1.25}
\begin{tabular}{lccc}
\toprule
\textbf{Method} &
\textbf{Power Loss} &
\textbf{VVR} &
\textbf{Max. Dev.} \\
&
\textbf{(MW)} &
\textbf{(\%)} &
\textbf{(p.u.)} \\
\midrule

Baseline 1 &
0.011027 &
2.824561 &
0.075099 \\

Baseline 2 &
0.005015 &
0.486842 &
0.057598 \\

Baseline 3 &
0.005399 &
0.405702 &
0.060522 \\

Baseline 4 &
0.005550 &
0.364035 &
0.056515 \\

\textbf{Proposed} &
\textbf{0.004407} &
\textbf{0.000000} &
\textbf{0.048881} \\

\bottomrule
\end{tabular}
\end{table}

To evaluate the autonomous performance of the trained policies, four baseline methods are considered. In addition to the previously considered baselines, Baseline~4 \cite{optimalVolt} is introduced as a rule-based correction module that modifies the action using predefined rules and performs only reactive power control, whereas the proposed HI-LSAC integrates an interactive human-guidance mechanism that coordinates both real and reactive power control across multiple devices for enhanced and safer voltage regulation.

For a fair comparison, all trained policies are evaluated using the same sequence of randomly selected load profiles. A total of 50 episodes, each consisting of 24 control intervals, are evaluated under stochastic policy execution using the same random seed for all methods. No online human correction is activated during evaluation. Therefore, the reported results characterize the autonomous control performance learned by each policy rather than the performance of an external safety intervention mechanism.

Table~\ref{tab:final_eval_powerloss_vvr} compares the autonomous evaluation performance of the trained policies in terms of active power loss, VVR, and maximum voltage deviation. Baseline~1 exhibits the poorest overall performance, with a power loss of $0.011027$~MW, a VVR of $2.824561\%$, and a maximum voltage deviation of $0.075099$~p.u. Although the policy is able to learn voltage-control actions, the relatively high VVR indicates inadequate voltage-security performance under stochastic policy execution.

Baseline~2 significantly improves the operating performance, reducing the power loss to $0.005015$~MW and the VVR to $0.486842\%$. However, the remaining voltage violations and a maximum voltage deviation of $0.057598$~p.u. indicate that conventional reward-based SAC learning alone is insufficient to consistently maintain the system within the prescribed voltage limits.

By explicitly incorporating voltage constraints into the learning process, Baseline~3 further reduces the VVR to $0.405702\%$. Nevertheless, its power loss increases to $0.005399$~MW, while the maximum voltage deviation reaches $0.060522$~p.u. These results demonstrate that adaptive Lagrangian learning improves overall voltage-security performance but does not necessarily provide the policy with explicit information regarding how an unsafe control action should be modified toward a safer alternative.

Baseline~4 incorporates human-guided corrective actions during training and achieves a lower VVR of $0.364035\%$ and a reduced maximum voltage deviation of $0.056515$~p.u. compared with Baseline~3. This improvement demonstrates that action-level human guidance provides additional learning information beyond the scalar reward and constraint signals. However, because Baseline~4 does not include adaptive Lagrangian constraint learning, its policy does not explicitly adjust the importance of voltage-security constraints according to the observed constraint violations during training.

The proposed HI-LSAC method achieves the best overall performance among all evaluated methods. It results in zero observed voltage violations, corresponding to a VVR of $0.000000\%$, while simultaneously achieving the lowest power loss of $0.004407$~MW and the smallest maximum voltage deviation of $0.048881$~p.u. The improvements over Baselines~3 and~4 demonstrate the complementary roles of adaptive Lagrangian constraint learning and human-guided corrective learning. The adaptive Lagrange multiplier continuously adjusts the influence of voltage violations during policy optimization, whereas the human-guidance mechanism provides direct action-level information when unsafe behavior is encountered. In addition, learning from rejected and corrected actions enables the agent to distinguish unsafe decisions from feasible corrective alternatives and progressively transfer this information to the learned policy.

\begin{figure}
    \centering
    \includegraphics[width=0.5\textwidth]{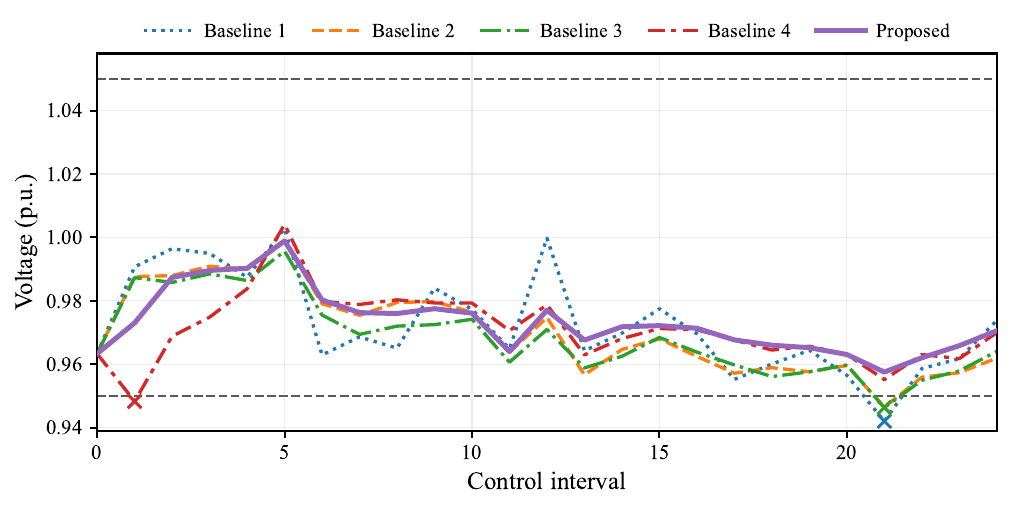}
    \caption{Evaluation-stage voltage trajectories at bus~680, phase~3, for all investigated methods under stochastic policy execution.}
    \label{fig:voltage_trajectory}
\end{figure}

Consequently, the proposed policy does not rely solely on constraint costs to infer safer behavior or solely on human corrections to reproduce individual intervention samples. Instead, the combined learning process enables the policy to internalize both the importance of voltage-security constraints and the corrective control behavior required to satisfy them. This complementary learning mechanism explains why HI-LSAC simultaneously achieves the lowest power loss and eliminates the voltage violations observed for all four baseline methods.

It should be emphasized that no human-guided correction is activated during the evaluation stage. Therefore, the zero observed VVR achieved by HI-LSAC is attributed to the learned policy itself rather than online action replacement, demonstrating that the corrective knowledge introduced during training is effectively transferred to autonomous policy execution.

Fig.~\ref{fig:voltage_trajectory} compares the evaluation-stage voltage trajectories at the critical bus~680, phase~3, for all investigated control methods under the same operating scenario. All baseline methods exhibit at least one voltage-limit violation during the episode, although Baseline~2 shows a less severe violation than the other baselines. In contrast, the proposed HI-LSAC method maintains the voltage within the prescribed $[0.95,\,1.05]$~p.u. range throughout the entire evaluation horizon.

\section{Conclusion}

This paper proposes HI-LSAC, a human-interactive safe RL framework for voltage control in unbalanced distribution networks. The proposed approach combines Lagrangian constraint learning with human-guided action correction to reduce unsafe exploration while improving the learning of effective voltage-control policies. The adaptive Lagrange multiplier dynamically regulates the influence of voltage constraints during training, while the human-guidance mechanism identifies unsafe actions and provides sensitivity-informed corrective actions that guide the policy toward feasible operating decisions. By incorporating these corrective interactions into the learning process, HI-LSAC enables the agent to progressively internalize safer control behavior rather than relying solely on online action correction. The case studies demonstrate that the proposed method achieves improved voltage-security performance while maintaining low network power losses compared with the benchmark DRL methods. These results highlight the potential of human-guided constrained RL as a practical framework for reliable and efficient voltage control in active distribution networks.

\ifCLASSOPTIONcaptionsoff
  \newpage
\fi



\bibliographystyle{IEEEtran}
\bibliography{mahmuda}

%







\end{document}